\documentclass[prb,showpacs,preprintnumbers,amsmath,amssymb,printfigures,twocolumn,superscriptaddress]{revtex4-1}
\usepackage[colorlinks]{hyperref}
\usepackage{graphicx}
\usepackage{amsmath}
\usepackage{bm}

\begin{document}
\title{Remarkable effect of stacking on the electronic and optical properties of few layer black phosphorus}

\author{Deniz \c{C}ak{\i}r}
\email{dcakir79@gmail.com}
\affiliation{Department of Physics, University of Antwerp, Groenenborgerlaan 171, B-2020, Antwerpen, Belgium}

\author{Cem Sevik}
\email{csevik@anadolu.edu.tr}
\affiliation{Department of Mechanical Engineering, Faculty of Engineering, Anadolu University, Eskisehir, TR 26555, Turkey}

\author{Francois M. Peeters}
\email{francois.peeters@uantwerpen.be}
\affiliation{Department of Physics, University of Antwerp, Groenenborgerlaan 171, B-2020, Antwerpen, Belgium}

\date{\today}

\begin{abstract}
The effect of the number of stacking layers and the type of stacking on the electronic and optical properties of bilayer and trilayer black phosphorus are investigated by using first principles calculations within the framework of density functional theory. We find that inclusion of many body effects (i.e., electron-electron and electron-hole interactions) modifies strongly both the electronic and optical properties of black phosphorus. While trilayer black phosphorus with a particular stacking type is found to be a metal by using semilocal functionals, it is predicted to have an electronic band gap of 0.82 eV when many-body effects are taken into account within the G$_0$W$_0$ scheme. Though different stacking types result in  similar energetics, the size of the band gap and the optical response of bilayer and trilayer phosphorene is very sensitive to the number of layers and the stacking type. Regardless of the number of layers and the type of stacking, bilayer and trilayer black phosphorus are direct band gap semiconductors whose band gaps vary within a range of 0.3 eV.  Stacking arrangments different from 
the ground state structure  in both bilayer and trilayer black phosphorus significantly modify valence bands along the zigzag direction and results in larger hole effective masses. The optical gap of bilayer (trilayer) black phosphorus varies by 0.4 (0.6) eV when changing the stacking type. Due to strong interlayer interaction, some stackings obstruct the observation of the optical excitation
of bound excitons within the quasi-particle band gap. In other stackings, the binding energy of bound excitons hardly changes with the type of stacking and is found to be 0.44 (0.30) eV for bilayer (trilayer) phosphorous.  
\end{abstract}

\maketitle

%optical properties, 78.67.-n :Semiconductors, III-V
%nanoscale materials, 73.22.-f
%optical properties of, 78.66.Db :Semiconductors, elemental
%Excitons, 71.35.-y
%Effective mass, 71.18.+y

\section{Introduction\label{intro}}

Triggered by the successful realization of a single layer of graphite (called graphene)\cite{gr1,gr2}, 
two dimensional (2D) layered materials have drawn a significant interest due to their potential use in the 
next generation of nanoelectronic  and optoelectronic  devices. The lack of a band gap in the 
electronic spectrum of graphene hampers its use in transistor applications, for instance\cite{zero-gap-1,zero-gap-2,zero-gap-3}. Due to its atomically thin structure and the electronic properties with a direct band gap of 1.8 eV, MoS$_2$ has been proposed as a promising candidate material\cite{mos2} for nanoelectronics. Field effect transistors (FETs) based on single layer of MoS$_2$ was reported to have an on/off ratio of 10$^8$ and a carrier mobility of 200 cm$^2$/V/s\cite{mos2-tran-1}, which may be further increased to 500 cm$^2$/V/s\cite{mos2-tran-2}. However, recent experiments have indicated that such a high carrier mobility in  monolayer MoS$_2$ may be just an overestimation due to the capacitive coupling between the gates of the devices\cite{mos2-tran-3}.

%mos2-tran-4

Very recently, a new direct band gap semiconducting 2D material, called black phosphorus, has been successfully exfoliated \cite{p1,p2,p4,chem-rev} and introduced to the 2D crystal family.  Besides its promising direct band gap, FETs based on black phosphorus have been reported to have a carrier mobility up to 1000 cm$^2$/V/s\cite{p1} and an on/off ratio 
up to 10$^4$ at room temperature\cite{p2}, making it a promising candidate material for future technological applications. Single layer BP has also been implemented into various electronic device applications including gas sensor~\cite{sensor}, \textit{p-n} junction~\cite{pn}, solar cell application~\cite{solar-cell} due to its sizable band gap ($\sim$ 0.9 eV) and because of its higher carrier mobility as compared to MoS$_2$\cite{p1,p2,bus,high-mob,BP-characater}, it is expected to be favorable for electronic device applications. 

Although the electronic\cite{qiao,dai,rodin,fei,vaitheeswaran,Zhu2014,Guan2014a} and optical\cite{bp-ex,bp-ex-2,dcakir,dcakir2,neto,Mehboudi} 
properties of mono and few layer phosphorenes have been investigated previously, the role of the stacking type on 
its electronic and optical properties have remained an open question so far. In the present study we investigate how its electronic and optical properties change with respect to the number of stacking layers and the stacking type. The paper is organized as follows: structural and electronic  properties of phosphorene is presented in Sec.~\ref{str-elect} and the optical response of monolayer phosphorene is presented in Sec.~\ref{optic}.  Our results are summarized in Sec.~\ref{conclusion}.

\section{Computational method}\label{method}
We carry out first-principles calculations in the framework of density functional theory (DFT) as implemented in the Vienna ab-initio simulation package (VASP)\cite{VASP,VASP1}. The generalized gradient approximation (GGA) within the Perdew- Burke-Ernzerhof (PBE)\cite{PBE} formalism and the Heyd-Scuseria-Ernzerhof (HSE06) hybrid functional\cite{hse-1,hse-2,hse-3} is employed for the exchange-correlation potential. The projector augmented wave (PAW) method\cite{paw} and a plane-wave basis set with an energy cutoff of 400 eV is used in the calculations. For geometry optimization the Brillouin-zone integration is performed using a regular 11 $\times$ 15 $\times$ 1 \textit{k}-mesh within the Monkhorst-Pack scheme\cite{monk}. The convergence criterion of the self-consistent field calculations is set to 10$^{-5}$ eV for the total energy. To prevent spurious interaction between isolated bilayers and trilayers, a vacuum spacing of at least 15 {\AA} is introduced. By using the conjugate gradient method, atomic positions and lattice constants are optimized until the Hellmann-Feynman forces are less than 0.01 eV/{\AA} and pressure on the supercells is decreased to values less than 1 kB. Since bilayer and trilayer black phosphorus are layered materials, we take into account the van der Waals interaction between individual layers for the correct description of the structural properties of black phosphorus\cite{grimme}.

The weak screening and reduced dimensionality in 2D systems
have a strong effect on their optical properties such that 
the excitonic effects have been shown to dominate the absorption spectra\cite{Louie,AVK-1,AVK-2,mose-ex}.
In such systems, many body interactions
(i.e., electron-electron and electron-hole interaction) must be taken
into account for a correct description of the optical properties. 
In this work, we investigate the effect of the number of layers and the type of stacking on the optical spectra  
in bilayer and trilayer black phosphorus using the Bethe-Salpeter-equation (BSE)
as implemented in the VASP code\cite{bse-1,bse-2}. 
In order to calculate the optical absorption spectrum and exciton binding energy, we follow the following steps.
First, hybrid-DFT calculations are performed by using PBE derived wavefunctions
within the HSE06 approach. This is followed 
by one-shot GW (i.e., G$_0$W$_0$) calculations to obtain the quasiparticle
excitations\cite{gw-1,gw-2,gw-3,gw-4,gw-5,gw-6}. Finally, we carry out BSE
calculations on top of G$_0$W$_0$ in order to obtain the optical adsorption spectra by
including excitonic effects using the Tamm-Dancoff 
approximation\cite{Tamm-Dancoff}.

The BSE calculations is performed on a 9$\times$13$\times$1 \emph{k}-mesh 
within the Monkhorst-Pack scheme. The energy cutoff for the 
wavefunctions and for the response functions is set to 400 eV and 200 eV, 
respectively. Since the number of empty bands significantly influences the 
relative position of the quasiparticle energy states, we use at least 100 empty bands. 
Within the current computational setup, the calculated quasiparticle gaps and exciton 
binding energies are then converged within 0.1 eV. 
The four highest occupied valence bands and four lowest unoccupied conduction bands is included as basis 
for the excitonic states. Since GW calculations require a sufficiently large 
vacuum region, we use a vacuum region of at least 15 \AA~ to avoid spurious 
interaction between the periodic images. A complex shift of $\eta$=0.1 eV is 
employed to broaden the calculated absorption spectra.

\section{Structural and electronic properties}\label{str-elect}
\begin{figure}
	\includegraphics[width=8cm]{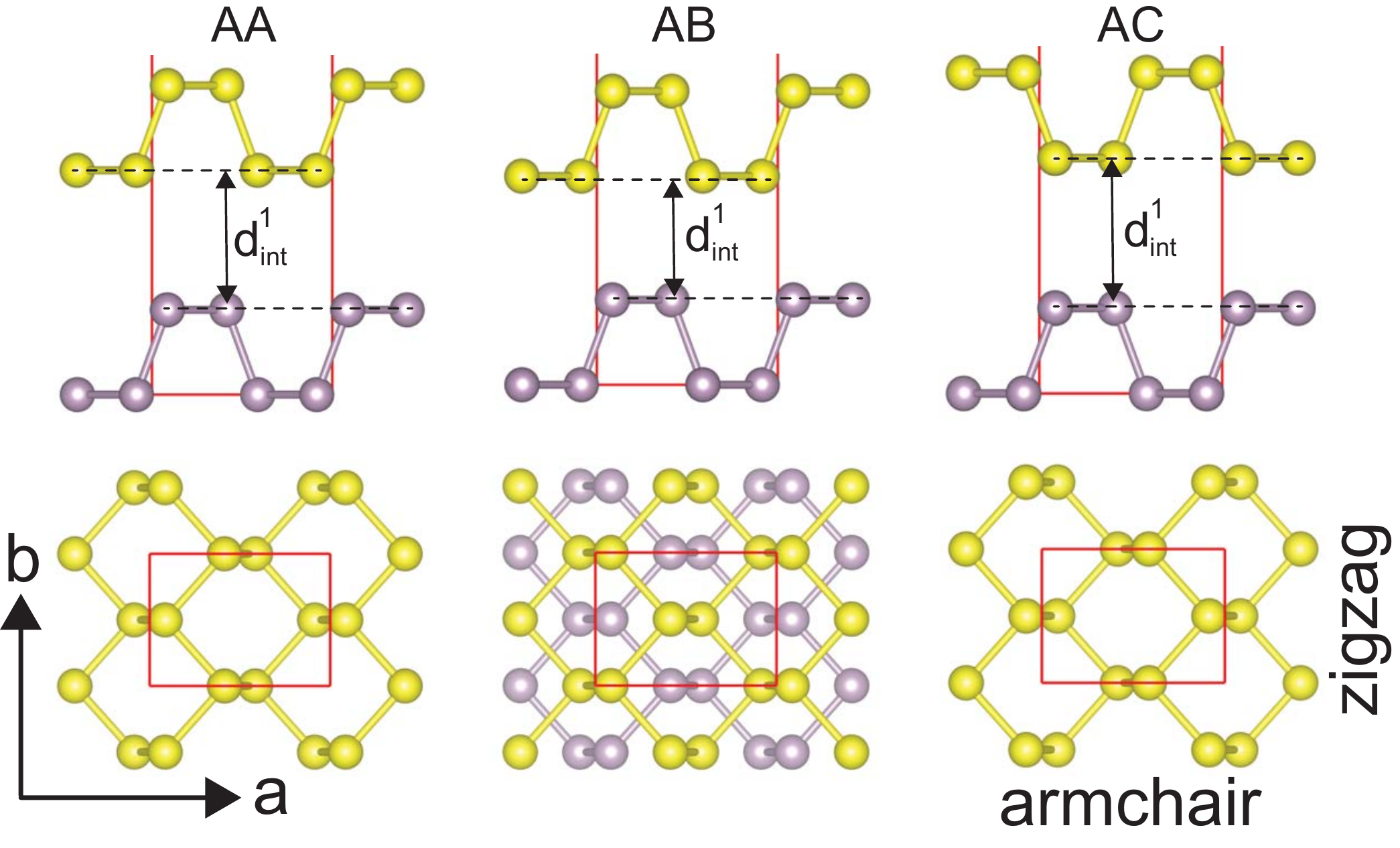}
	\caption{(Color online) Side and top views of AA, AB and AC stacking structures for bilayer black phosphorus. $d^1_{int}$  denotes the interlayer distance. We also show the structural definition of the zigzag and armchair directions of black phosphorus.}
	\label{fig:2L-str}
\end{figure}

\begin{figure}
	\includegraphics[width=8cm]{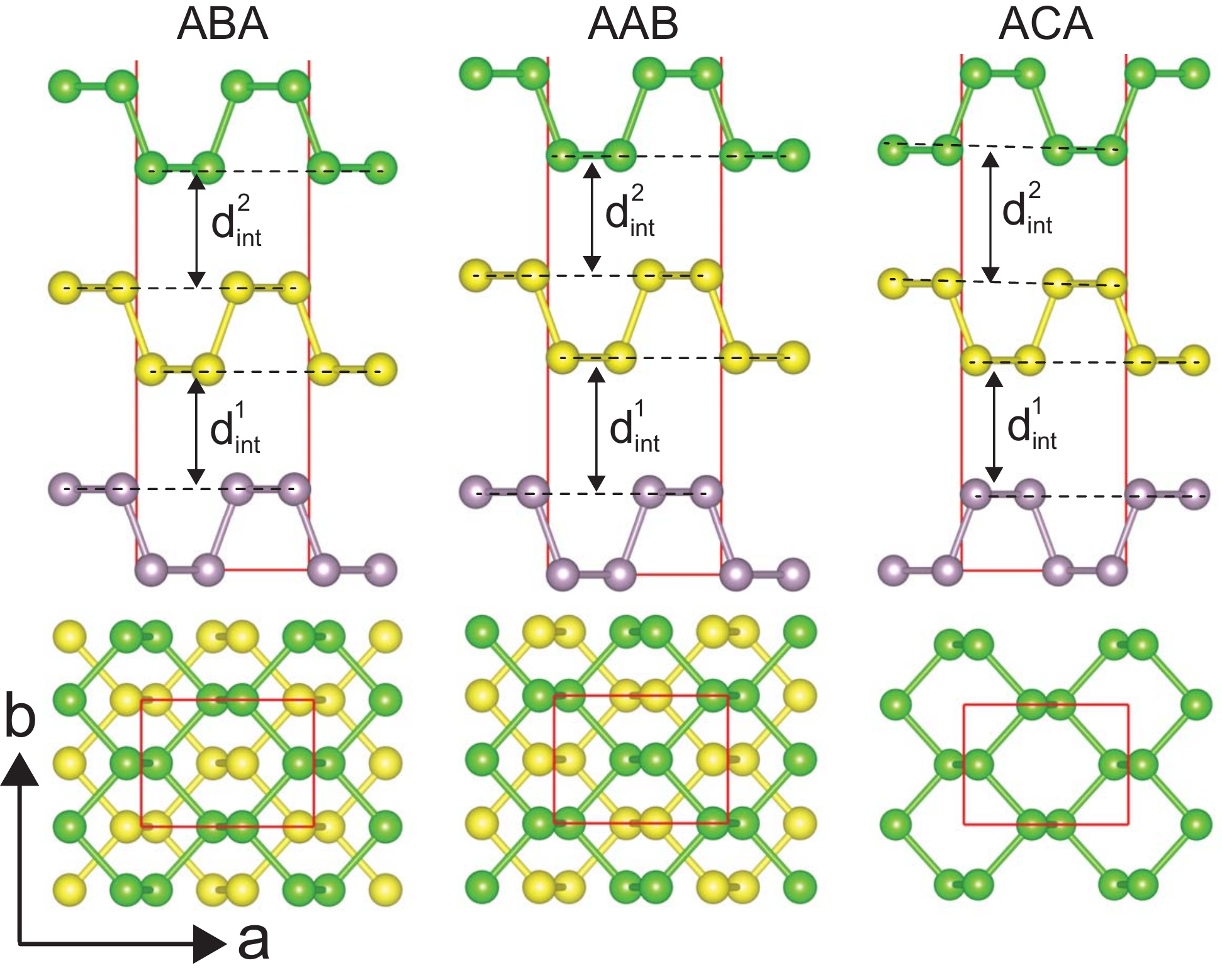}
	\caption{(Color online) Side and top views of ABA, AAB and ACA stacking structures for trilayer black phosphorus. $d^1_{int}$ and $d^1_{int}$ denote the interlayer distances.}
       \label{fig:3L-str}
\end{figure}

In this study, three possible stacking types, illustrated in Figs.~\ref{fig:2L-str} and \ref{fig:3L-str}, are considered for both bilayer and trilayer black phosphorus. To realize these stacking types, individual monolayers were shifted with respect to the ground state structure. For instance, the top layer of AB stacking in bilayer is shifted by half of the cell along  either the $a$ or $b$ direction in order to obtain AA stacking. The calculated structural parameters (\textit{a} and \textit{b}) summarized in Table~\ref{table1} differ slightly for different stacking types. The most notable difference among the different stacking types is the interlayer distance between individual layers. It varies from 3.11 {\AA} in AB stacking to 3.70 {\AA} in AC stacking. In AAB stacking of trilayer, we have two quite different interlayer distances, namely 3.11 {\AA} between bottom and middle layers and 3.51 {\AA} between middle and top layers.

To reveal the effect of the stacking type on the energy, we calculate the total energy difference $\Delta E$ (per atom) between a particular stacking type and the ground state structure. As is evident from Table~\ref{table1}, while stacking types are quite different in both bilayer and trilayer systems, $\Delta E$ is found to be on the order of 20 meV/atom, meaning that at finite temperatures (for instance at room temperature) one may have a high energy structure, for instance AC stacking type in bilayer and ACA stacking type in the trilayer case. This highlights the importance of studying the impact of the number of stacking layers and stacking type on the electronic and optical properties of black phosphorus. 

\begin{figure}
	\includegraphics[width=8cm]{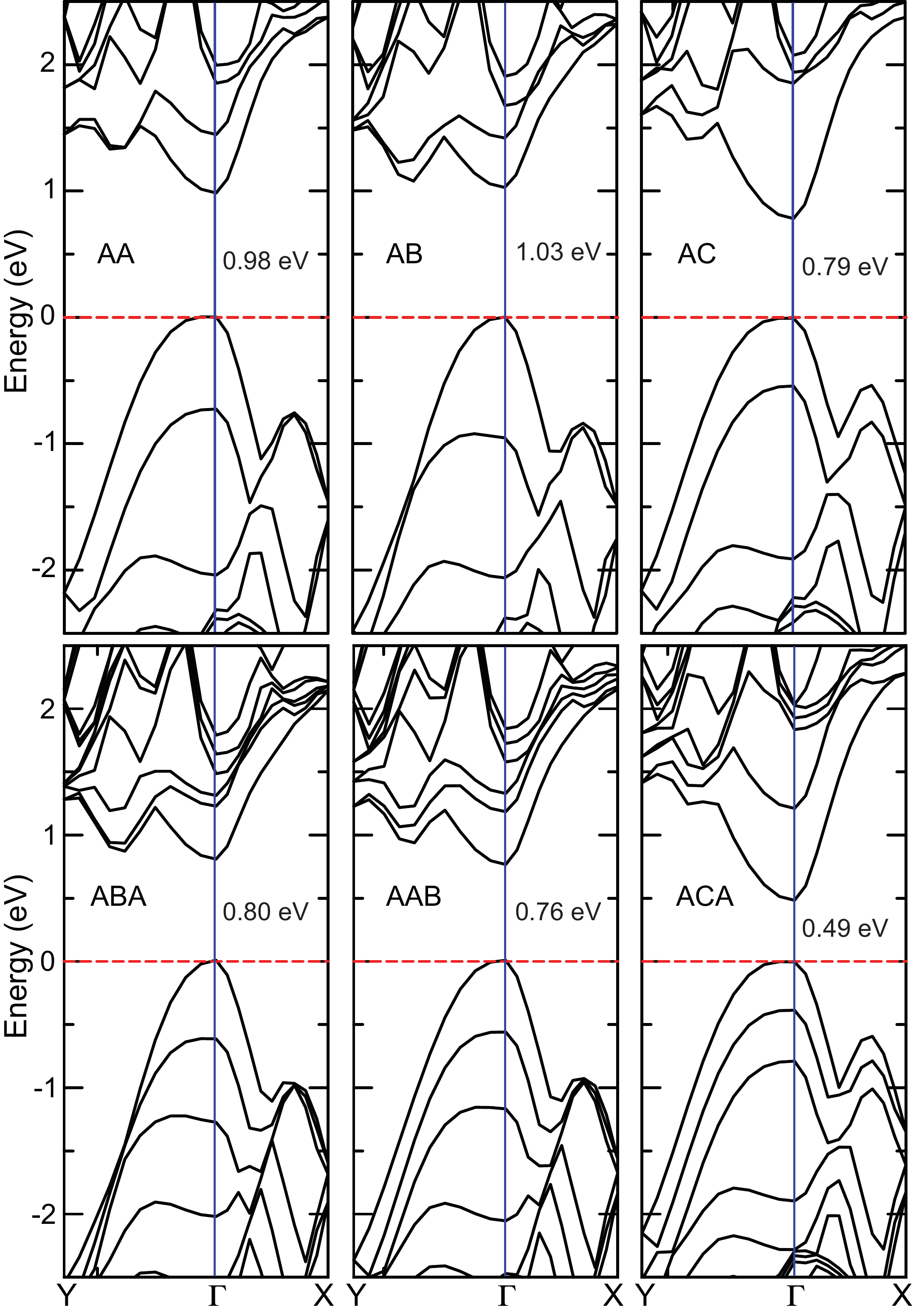}
	\caption{(Color online) Band structures computed with HSE06 along the Y-$\Gamma$-X direction. The band gap values are also given for each structure. The Fermi level is set at 0 eV. }
       \label{fig:band-str}
\end{figure}

Table~\ref{table1} summarizes the electronic band gap ($E_{gap}$) values calculated with PBE, HSE06 and G$_0$W$_0$ for bilayer and trilayer black phosphorus for different stacking types. As seen in Table~\ref{table1} using hybrid functionals significantly enlarges $E_{gap}$. For instance, while $E_{gap}$ for bilayer black phosphorus with AB stacking is calculated as 0.42 eV with PBE-GGA, it becomes 1.03 eV when using HSE06, consistent with previous calculations\cite{dai}. Figure~\ref{fig:band-str} shows the band structures calculated using the HSE06 functional for all structures.  Importantly, inclusion of many body effects (G$_0$W$_0$) further increases $E_{gap}$ to 1.45  eV (see Table~\ref{table1}).  Trilayer ACA  stacking is found to display metallic behavior when using the PBE functional. However, both HSE06 and G$_0$W$_0$ predict that it is a direct band gap semiconductor with an electronic band gap of 0.49 (0.82) eV with HSE06 (G$_0$W$_0$). Therefore, to predict the electronic ground state accurately, hybrid and GW calculations are essential. We observe that the electronic band gap of bilayer and trilayer black phosphorus can vary by 0.3 eV when changing the stacking type, consistent with a recent work based on first-principles calculations\cite{dai}. For instance, AA, AB and AC-stacked bilayer black phosphorus have a quasi-particle gap of 1.40, 1.45 and 1.24 eV, respectively. Since $\Delta E$ is quite small for a particular stacking, it is expected that  mixed stacking in multilayer black phosphorus are found in experiments. In addition, this competing energetics among different stacking types makes it easier to tune the electronic properties. Another important point is that, regardless of the number of layers and stacking type, bilayer and trilayer black phosphorus have a direct band gap at the $\Gamma$ point.

\begin{figure}
	\includegraphics[width=8.5cm]{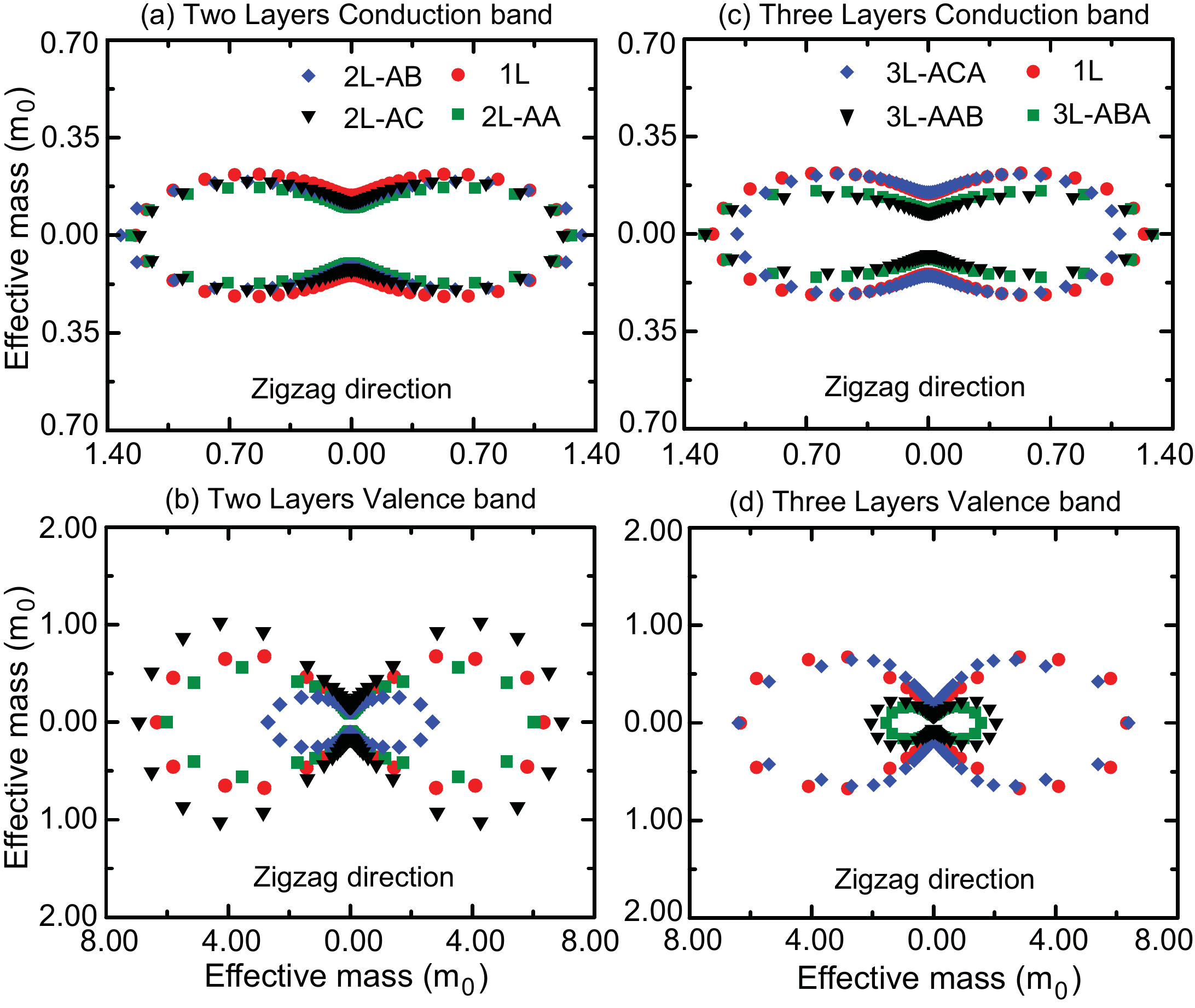}
	\caption{(Color online) Direction dependent hole and electron effective masses for the different  number of layers and different types of stacking. $x$ ($y$) axis denotes the zigzag (armchair) direction of black phosphorus. Each data point represents the end point of a vector whose amplitude corresponds to the effective mass in units of m$_0$.}
	\label{fig:effective-mass}
\end{figure}

\begin{table*}
\caption{\label{table1} Calculated interlayer distance ($d^1_{int}$ and $d^2_{int}$), lattice parameters ($a$ and $b$),  
electronic band gaps ($E_{gap}$) for different exchange-correlation functionals, optical gaps 
($E_{opt}$) and exciton binding energies ($E_{bind}$= $E_{gap}^{G_0W_0}$-$E_{opt}$) for bilayer and trilayer black phosphorus for different stacking types. Distances and lattice parameters are given in units of {\AA}. Band gap values and energies are given in units of eV. $\Delta E$ is the total energy difference (in units of meV/atom) between a particular stacking type and the ground state structure calculated with the PBE functional including van der Waals correction. Here, the AB (ABA) stacking represents the ground state structure in bilayer (trilayer) black phosphorus. Electron ($m^e_z$ and $m^e_a$) and hole ($m^h_z$ and $m^h_a$) effective masses at the $\Gamma$ point along the zigzag ($z$) and armchair ($a$) directions are given in units of the free electron effective mass (m$_0$). For comparison, we also include the results for  single layer (1L) black phosphorus. }
\begin{tabular}{lccccccccccccccc}
\hline\hline
  &$d^1_{int}$&$d^2_{int}$& $a$ & $b$ & $E_{gap}^{PBE}$ & $E_{gap}^{HSE06}$ & $E_{gap}^{G_0W_0}$ & $E_{opt}$& $E_{bind}$ & $\Delta E$ & $m^h_z$ & $m^e_z$ & $m^h_a$ & $m^e_a$  \\
\hline
1L &        &  & 4.61     & 3.30      & 0.90&1.59&2.31&1.61&0.7&       & 6.33&1.24&0.13&0.14 \\
AA& 3.50& &4.51&3.31&0.38&0.98&1.40&0.96&0.44&13.0&   6.02  & 1.26    &  0.09   &  0.10   \\
AB& 3.09& &4.51&3.31&0.42&1.03&1.45&1.02&0.43&0.0&   2.70  & 1.32    &  0.11   &  0.11    \\
AC& 3.71& &4.52&3.31&0.20&0.79&1.24&1.37&0.46&20.0&   6.93  & 1.22    &  0.17   &  0.12  \\
ABA&3.13 &3.13&4.50&3.32&0.21&0.80&1.10&0.81&0.29&0.0&  1.55  & 1.29    &  0.08   &  0.09     \\
AAB&3.51 &3.11&4.50&3.31&0.19&0.76&1.06&0.76&0.30&8.6&  2.05  & 1.29    &  0.07   &  0.08    \\
ACA&3.68 &3.69&4.53&3.30&0.00&0.49&0.82&1.34&0.33&27.0& 6.39 & 1.10    & 0.19  & 0.15      \\
  \hline \hline
\end{tabular}
\end{table*}

Besides having a direct band gap at the $\Gamma$ point, black phosphorus has a highly anisotropic band structure around the band gap. Both valence  and conduction bands have a much more significant band dispersion along the armchair direction.  Table~\ref{table1} summarizes hole and electron effective masses both along the zigzag and the armchair directions at the $\Gamma$ point. Since the effective mass is directly proportional to the inverse of the curvature of the band dispersion, both electrons and holes have  much larger effective masses along the zigzag direction. For different directions, the effective masses can even differ by an order of magnitude. For instance, while the electron effective mass is calculated 0.10 m$_0$ along the armchair direction, it becomes 1.26 m$_0$ along the zigzag direction for the AB stacking of bilayer black phosphorus. Here
m$_0$ denotes the effective mass of a free electron.  It is worth revealing the dependence of the effective masses on both the number of layers and the stacking type. The spatial variation of the  effective mass of holes and electrons of monolayer, bilayer and trilayer black phosphorus is illustrated in Fig.~\ref{fig:effective-mass}. Each data point represents the end point of a vector that shows the effective mass calculated along the corresponding reciprocal space direction. The first observation is that their variation has approximately an '8' shape, which proves the anisotropic nature of the electronic properties of black phosphorus. In addition, the variation of the effective hole mass is quite different as compared to that of the effective electron mass. Consistent with previous work\cite{qiao}, the effective mass of the hole along the zigzag direction decreases with increasing the number of layers for the lowest energy stacking types i.e., AB and ABA. For the armchair direction, no appreciable variation is observed. However, the type of stacking may have a significant impact on the effective masses. As one changes the stacking from ABA to AAB in trilayer black phosphorus, the hole effective mass along the zigzag direction increases from 1.55 m$_0$ to 6.39 m$_0$. Such a large change will strongly reduce  the flow of holes along the zigzag direction.  These  anisotropic electronic properties force to confine the carriers in an effective one-dimensional environment along the armchair direction. Since both bilayer and trilayer black phosphorus for AA, AC, AAB and AAC stackings have a rather flat band structure for the valence band along the zigzag direction, the hole effective mass is quite large for these type of stackings, see Figs.~\ref{fig:band-str} and \ref{fig:effective-mass}.  
Note that departing from the lowest energy stacking type retains its monolayer behavior for the valence band  along the zigzag direction in both bilayer and trilayer black phosphorus. 
However, the electron effective mass is insensitive to the stacking along the zigzag direction. Interestingly, both electron and hole effective masses along the armchair direction are about an order of magnitude smaller than those along the zigzag direction and are hardly affected by the type of stacking. Our calculations suggest that an appropriate control of the type of stacking  may help to control especially hole conduction in phosphorene based electronic devices. 

\begin{figure*}
	\includegraphics[width=15cm]{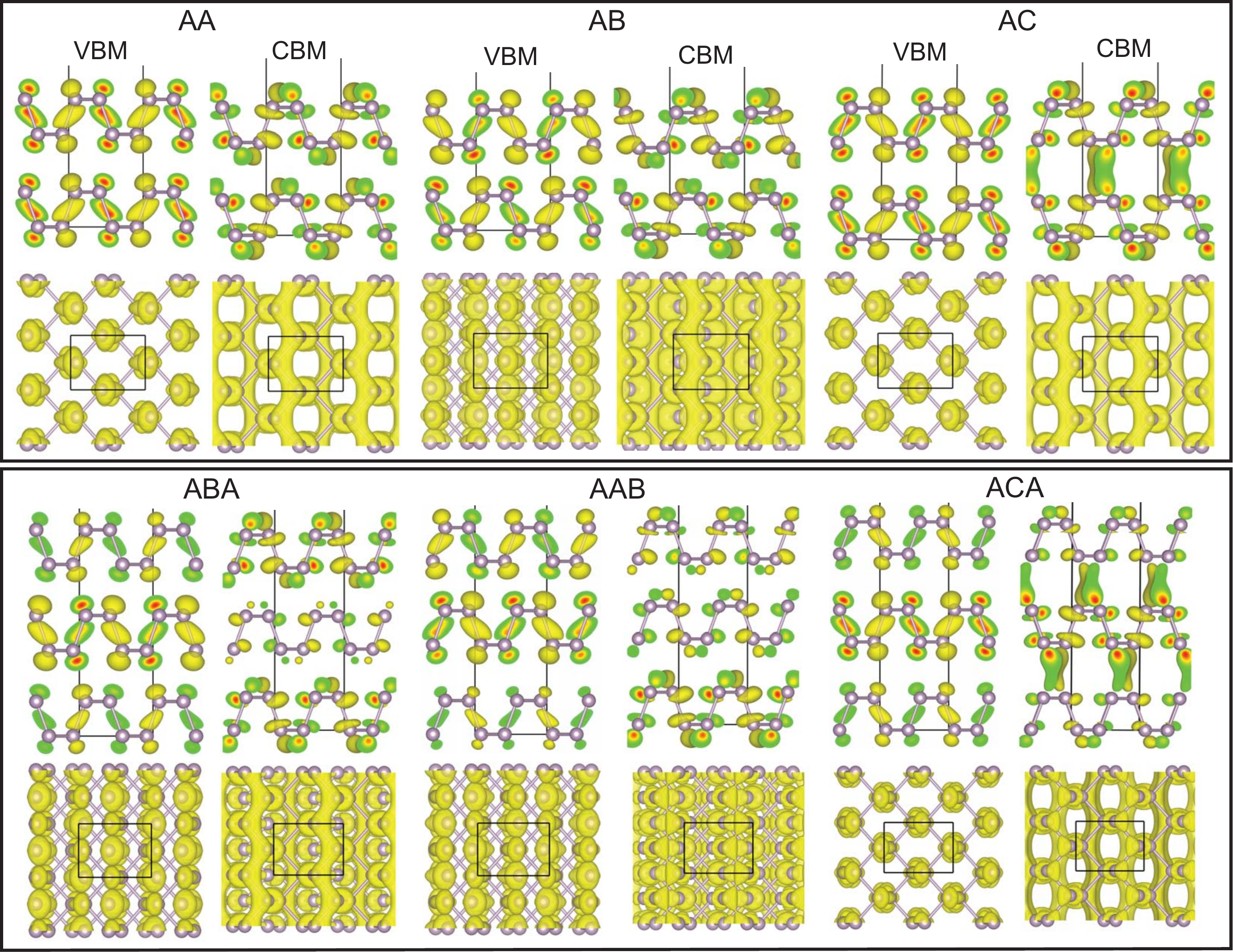}
	\caption{(Color online) Side and top views of decomposed charge densities corresponding to the VBM and CBM for both bilayer and trilayer black phosphorus at the $\Gamma$ point.}
	\label{fig:chg-den}
\end{figure*}

Figure~\ref{fig:chg-den} shows the band decomposed charge densities corresponding to the valence band maximum (VBM) and the conduction band minimum (CBM) at the $\Gamma$ point. In all cases, the CBM has a delocalized nature along the zigzag direction of black phosphorus. Stacking especially modifies  the CBM in AC stacking in bilayer and ACA in trilayer. We notice partially delocalized states in the interfacial area between layers, which can be associated with a larger band gap lowering in AC and ACA stacked black phosphorus as compared to other stacking types. The VBM charge distribution is relatively insensitive to the stacking type and has a localized nature. Due to the presence of different interlayer interactions in the trilayer case,  the contribution of individual layers to the charge density at the VBM varies. As is clear from Fig.~\ref{fig:chg-den} different stacking types results in different $\pi$-$\pi$ interaction distances and strengths, which are responsible for the tuning of the electronic band gaps.

\section{Optical properties}\label{optic}
The optical gaps ($E_{opt}$) and exciton binding energies ($E_{bind}$ = 
$E^{G_0W_0}_{gap}$ - $E_{opt}$) are listed in Table~\ref{table1}.
Though the stacking types have a small effect on the calculated total energy and lattice parameters,
we find that not only $E_{gap}$ but also $E_{opt}$ and $E_{bind}$ are sensitive to the type of stacking. 
This means that the electronic and  optical properties of black 
phosphorus can be modified by changing the type of stacking\cite{dai}. 
Before discussing the optical properties of  bilayer and trilayer black phosphorus, 
it is informative to summarize the results for the single layer case. 
The quasi-particle band gap obtained from G$_0$W$_0$ is 2.3 eV.
In line with our work,  photoluminescence excitation spectroscopy 
find a quasi-particle band gap of 2.2 eV\cite{bp-ex-3}.
Such a large modification of PBE-GGA band gap upon inclusion of many body 
effects is a result from the enhanced electron-electron correlation 
due to confinement effect in 2D black phosphorus.
Similarly, the experimental  optical gap\cite{p2} for single layer black phosphorus 
was measured to be around 1.45 eV in Ref.~\citenum{p1} and 1.3 eV in Ref.~\citenum{bp-ex-3}, 
in agreement with our calculated value of 1.60 eV\cite{dcakir}.  
Due to the reduced dimensionality and weak screening, 
single layer black phosphorus has a large exciton binding energy of 0.7 eV, 
which is in good agreement with recent theoretical works.\cite{bp-ex,bp-ex-2}
By applying a tensile strain of 4\%, $E_{bind}$ can be further increased to 0.83 eV, 
placing it among the highest of the 2D materials\cite{dcakir}. 
The optical properties of single layer black phosphorus shows a strong orientation 
dependence such that the wavefunction of bound excitons are extended along the armchair direction. 

\begin{figure}
	\includegraphics[width=8.3cm]{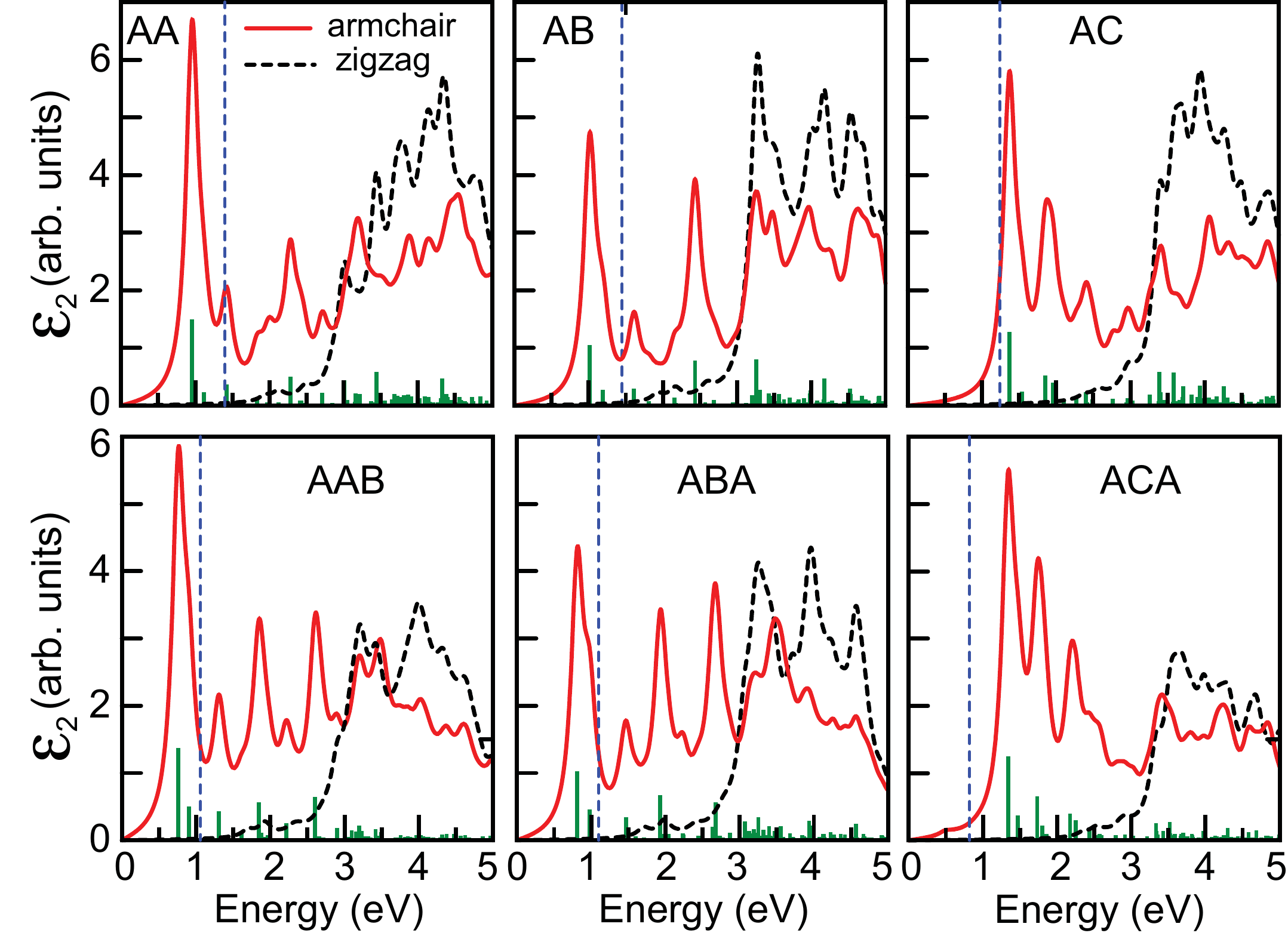}
	\caption{(Color online) G$_0$W$_0$+BSE absorption spectra for bilayer and trilayer black phosphorus with different stacking types. Blue vertical dashed lines mark the electronic band gap calculated at the level of G$_0$W$_0$. Green vertical lines represent the relative oscillator strengths for the different optical transitions.}
	\label{fig:bse}
\end{figure}

%the main optical features are dominated by excitonic states
%the prominent absorption features begin at higher energies for zigzag
%this phenomenon is the result of selection rules associated with the symmetries of this anisotropic material.

Figure~\ref{fig:bse} displays the optical absorption spectra ($\varepsilon_2$) of 
bilayer and trilayer black phosphorus having different stacking types for light polarized 
along the armchair and zigzag directions. We also show the electronic 
band gap values calculated with G$_0$W$_0$ in the same plots. Similar to single layer black phosphorus,  
the optical absorption spectra displays a strong orientation dependence.
Due to the reduced dimensionality and weak screening,  excitonic effects largely shape the optical spectra.
A strong absorption peak along the armchair direction is located around 0.96 eV for AA, 1.02 eV for AB,
0.81 eV for ABA  and 0.76 eV for AAB, which are bound excitonic states with an $E_{bind}$ of 0.44 eV for bilayer and 0.30 eV for trilayer phosphorus. 
All these peaks originate from the interband transition from VBM to CBM. 
However, AC and ACA stackings behave differently. In spite of their smaller electronic band gaps as compared to
other stacking types, we observe a strong absorption peak  around 1.37 eV for AC stacking in bilayer and 1.34 eV for ACA stacking in trilayer, which are located
0.13 eV for the former and 0.52 eV for the latter above the quasi-particle band gap. 
Actually, there is a very small absorption feature at 0.79 eV for AC stacking and 0.49 eV for ACA stacking with a very small oscillator strength, corresponding to an excitonic state with a binding energy of 0.46 (0.33) eV in former (latter) stacking.  These optical transitions result in a very small intensity in a photoluminescence experiment. So, their observation is expected to be difficult.  
This weak intensity may be linked to the different behavior of the CBM at the $\Gamma$ point. While the formation of AC and ACA are energetically less favorable as compared to other stacking types, the CBM has a delocalized nature in the interfacial region, which further enhances the interlayer interaction, thereby preventing the excitation of a bound exciton. The optical response of bilayer (trilayer) black phosphorus can be tuned from 0.96 (0.76) eV for AA (AAB) to 1.37 (1.34) eV for AC (ACA). Such a large change in the optical absorption energy window can be utilized in the design of efficient photovoltaic and optoelectronic devices.  

As expected, $E_{bind}$ decreases with increasing number of stacking layers  due to the enhancement of the interlayer interaction. In Ref.~\citenum{bp-ex}, $E_{bind}$ has been found to be only 30  meV for an infinite number of layers i.e., bulk black phosphorus. We observe that bound exciton wavefunctions are mainly extended along the armchair direction. Adversely, there is no optically bright exciton whose wavefunction is extended along the zigzag direction. This means that BP is transparent to polarized light along the zigzag direction and  when an exciton is optically excited, it forms along the armchair direction.
Due to selection rules associated with the symmetries of black phosphorus, the prominent absorption peaks along the zigzag direction start at energies about 2 eV above the quasi-particle band gap.
These peculiar optical properties make black phosphorus a promising  optical linear polarizer, which can be used in several applications, for instance in liquid-crystal displays.

%While we did not explicitly calculate the exciton wavefunctions, Fig.~\ref{fig:chg-den} provides some clues about how the exciton wavefunction extends within the trilayer, for instance. In ABA stacking, VBM has a dominant %contribution coming from the middle layer. In contrast, the charge density of CBM is mainly distributed on the bottom and the top layers. In Ref.~\citenum{bp-ex}, it has been found that the wave function of the hole is fixed to one %layer only and the electron is distributed over two layers, consistent with Fig.~\ref{fig:chg-den}.

Lastly, we discuss the impact of the interlayer distance on the exciton binding energy. While AA and AB stackings have quite different interlayer distances, $E_{bind}$ is found to be around 0.44 eV for both types of stacking. Similarly, in the trilayer case, in spite of different interlayer separations in ABA and AAB stackings, $E_{bind}$ converges to an average value of 0.30 eV.  Due to the vacuum surrounding these stacked layers, $E_{bind}$ hardly changes with interlayer separation. 

\section{Conclusion}\label{conclusion}

In summary, our results show that the number of layers and the type of stacking have
a significant impact on the electronic and optical properties of black phosphorus. 
The change of the band gap with stacking is directly related with the downward shift of the CBM, 
originating from the presence of different interlayer interactions. 
However, the nature of the band gap (i.e., direct band gap at the $\Gamma$ point) is insensitive to the number of layers and the type of stacking.  
The hole effective mass strongly depends on the type of stacking. While the hole effective 
mass at the $\Gamma$ point along the zigzag direction decreases 
with increasing number of layers, the deviation from 
the lowest energy structures results in  quite flat valence bands, thereby increasing the hole effective masses along the zigzag direction. 
We find that the absorption energy window of black phosphorus can be tuned by changing the 
number of layers and by the type of stacking. While stacking alters the optical gap of bilayer black phosphorus 
up to 0.4 eV, it may result in a 0.6 eV change in the trilayer case.
Previously strain engineering has been shown to modify the electronic and optic properties of black phosphorus\cite{dcakir}. 
Here, we show that playing with the number of layers and the type of stacking appears as an exciting alternative way to tune the optical response and the electrical properties of black phosphorus.
Finally,  it is worth mentioning that the anisotropic optical and electronic properties can be utilized to distinguish the type of stacking and  the orientations of few-layer black phosphorous in experiments.

\section{ACKNOWLEDGMENTS}
This work was supported by the Flemish Science Foundation (FWO-Vl) and the Methusalem foundation of the Flemish government. Computational resources were provided by TUBITAK ULAKBIM, High Performance and Grid Computing Center (TR-Grid e-Infrastructure), and HPC infrastructure of the University of Antwerp (CalcUA) a division of the Flemish Supercomputer Center (VSC), which is funded by the Hercules foundation. C.S. acknowledges the support from Turkish Academy of Sciences (TUBA-GEBIP).

\bibliography{ref}
\end{document}